\documentclass[letter, twocolumn]{jpsj3}
\usepackage{txfonts}
\usepackage{braket}
\usepackage{bm}

\usepackage{graphicx}
\usepackage{dcolumn}
\usepackage{bm}
\usepackage{amsmath}
\usepackage{times}
\usepackage[usenames]{color}

\title{Photo-Induced Dynamics in Charge-Frustrated Systems}

\author{Hiroshi Hashimoto$^1$\thanks{hashimoto@cmpt.phys.tohoku.ac.jp}, Hiroaki Matsueda$^2$, Hitoshi Seo$^{3,4}$, and Sumio Ishihara$^{1,5}$}
\inst{$^1$Department of Physics, Tohoku University, Sendai 980-8578, Japan \\
$^2$Sendai National College of Technology, Sendai 989-3128, Japan \\
$^3$Condensed Matter Theory Laboratory, RIKEN, Wako, Saitama 351-0198, Japan \\
$^4$RIKEN Center for Emergent Matter Science (CEMS), Wako, Saitama 351-0198, Japan \\
$^5$JST-CREST, Sendai 980-8578, Japan} 

\abst{
Photo-excited charge dynamics of interacting charge-frustrated systems are studied using a spinless fermion model on an anisotropic triangular lattice.
Real-time evolution of the system
after irradiating a pump-photon pulse is analyzed by the exact diagonalization method. 
We focus on photo-excited states in the two canonical charge-ordered (CO) ground states,
i.e., horizontal stripe-type and vertical stripe-type COs,
which compete with each other owing to the charge frustration. 
We find that the photo-induced excited states from the two types of COs are distinct. 
From the horizontal stripe-type CO,
a transition to another CO state called the three-fold CO phase occurs.
In sharp contrast, the vertical stripe-type CO phase is only weakened by photo-irradiation. 
Our observations are attributable to the charge frustration effects occurring in the photo-excited states.  
}

\kword{charge order, photo-induced phase transition, frustration}

\begin{document}
\maketitle


Frustration effect is ubiquitously seen in a wide field of phenomena in
condensed matter physics~\cite{JPSJ_frustration,book_frustration}. 
Situations where a macroscopic number of degeneracy exists in the
classical ground state
and where exotic quantum states are caused by  releasing the degeneracy have attracted much attention experimentally and theoretically.
Well studied examples are interacting spin systems in geometrically frustrated lattices, where quantum spin-liquid, spin-ice, spin-nematic states, multiferroics, and so on, have been realized~\cite{shimizu,han,harris,nakatsuji,kimura}. 
Another class of frustration effect has been attracting interest as well: in interacting charge systems, i.e. charge frustration. 
This concept was first proposed by Anderson~\cite{anderson} for the Verwey transition in magnetites where equal amount of  Fe$^{2+}$ and Fe$^{3+}$  occupy the so-called B sites in the spinel crystal lattice~\cite{verwey}. 
Similar situation of charge ordering (CO) phenomena in frustrated lattices is realized in several transition-metal oxides as well as organic molecular materials~\cite{seo,merino,hotta,nishimoto0}. 

Then, a question arises: how does the frustration effect emerges in highly excited states and non-equilibrium states. 
This question can be explored, in particular, by the optical pump-probe experiments in interacting CO systems. 
A concept of frustration is not only applicable to the low-lying excited states,  
but also is conceivable to induce a large degeneracy in optically generated highly excited states. 
An expected consequence is the generation of a multi-valley energy surface which should crucially govern photo-induced charge dynamics after optical excitation as well as in its relaxation. 
An optically induced hidden CO state, which is not realized in the thermal equilibrium state, would be an another example 
driven by charge frustration effect in highly excited states. 
In recent decades, time-resolved experimental techniques~\cite{nasu,JPSJ_PIPT,chollet,iwai,rohwer,stojchevska} as well as theoretical calculations~\cite{yonemitsu,iwano,kanamori,gomi} 
have been applied  to photo-induced phenomena in CO systems. 
Now we are in the stage of having a chance to explore the frustration concept in optically induced excited and non-equilibrium states in interacting electron systems. 

\begin{figure}[t]
\includegraphics[width=\columnwidth,clip]{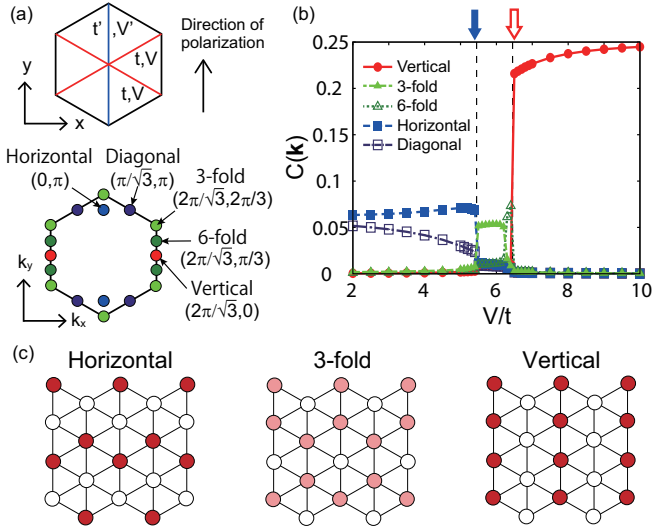}
\caption{
(Color online) 
(a) (upper panel) Fermion hoppings and Coulomb interactions on the triangular lattice. 
(lower panel) The first Brillouin zone and momenta corresponding to different types of COs.   
Momenta for the v-stripe, three-fold, six-fold, h-stripe, and d-stripe COs  
are $(2\pi/\sqrt{3}, 0)$, $(2\pi/\sqrt{3}, 2\pi/3)$, $(2\pi/\sqrt{3},\pi/3)$, $(0, \pi)$ and $(\pi/\sqrt{3},\pi)$, respectively. 
(b) The $V$-dependence of charge correlation functions in the ground state keeping $V+V'=12t$. 
Filled and open arrows 
indicate parameter values chosen in Fig.~\ref{fig:ckt}(a) and (c), respectively. 
(c) Schematic representations for the h-stripe, v-stripe and three-fold CO structures. 
Filled and open circles represent charge rich and poor sites, respectively. }
\label{fig:gs}
\end{figure}

In this Letter, we examine photo-excited states in a typical CO system on a geometrically frustrated lattice. 
As a minimum model to explore such charge frustration effects in highly excited states, 
we adopt an interacting spinless fermion model on a triangular lattice, where a number of fermion per site is 1/2. 
This model has been studied in terms of the CO phenomena in organic molecular materials, 
such as $\alpha$-type and $\theta$-type BEDT-TTF compounds~\cite{hotta,naka,nishimoto}. 
The ground-state phase diagram is obtained on the plane of the nearest-neighbor (NN) Coulomb-interactions
on the anisotropic triangular-lattice bonds, i.e. $V$ and $V'$ (see Fig.~\ref{fig:gs}(a)). 
Two canonical insulating CO states, called horizontal stripe-type
(abbreviated as h-stripe) and vertical stripe-type (v-stripe) COs, are realized
in the classical limit of weak fermion hopping, for $V < V'$ and $V > V'$, respectively, and the two phases compete with each other around $V =V'$. 
When a finite fermion hopping is included, a three-fold CO metallic phase appears between the two COs. 
We examine numerically the photo-excited real-time dynamics of this system
by using the exact diagonalization technique. 
It is found that the photo-excited dynamics in the two insulating COs are distinct; 
a photo-induced transition from the h-stripe 
to the three-fold CO occurs, whereas, on the other hand, the CO amplitude 
is only weakened in the v-stripe CO phase.
This difference is attributable to the frustration effects 
in the photo-excited states, where a large number of low-lying states exists.

The model we study is given by the Hamiltonian:
~\cite{hotta,nishimoto0,nishimoto} 
\begin{align}
 {\cal H}= -\sum_{\braket{ij}}t_{ij} c_{i}^{\dagger}c_{j} 
+ \sum_{\braket{ij}} V_{ij} n_{i}n_{j} , 
\label{e1}
\end{align}
where $c_{i}$ ($c_{i}^{\dagger}$) is an annihilation (creation) operator for a spinless fermion at site $i$ 
and $n_{i}=c_{i}^{\dagger}c_{i}$ is a number operator.
The first and second terms represent the fermion hoppings and the inter-site 
Coulomb interaction, respectively, between the NN sites on a triangular lattice.
As shown in Fig.~\ref{fig:gs}(a), anisotropies in the fermion hopping and the Coulomb interactions are represented as $(t, t')$, and $(V, V')$, respectively. 
The optical pump pulse is introduced as the Peierls phase into the transfer integral as 
$t_{ij}\rightarrow t_{ij}e^{-{\rm i}\bm{A}(\tau)\cdot \bm{R}_{ij}}$
where ${\bm A}(\tau)$ is the vector potential at time $\tau$, and $\bm{R}_{ij}$ is a relative position vector connecting sites $i$ and $j$.
The time dependence of the vector potential is assumed to be a Gaussian form given by 
$\bm{A}(\tau) = A_{p}\bm{e}(\sqrt{2\pi}\tau_{p})^{-1}
e^{-\tau^{2}/2\tau_{p}^{2}} \cos(\omega_{p}\tau)$ 
with amplitude $A_{p}$, frequency $\omega_{p}$, a damping factor $\tau_p$, 
and a unit vector $\bm{e}$. 

The ground state and the photo-excited transient states are calculated by using the exact diagonalization method based on the Lanczos algorithm.  
We take two-dimensional finite-size clusters up to $N=4\times 6=24$
sites  with the periodic boundary condition; in the following we show
the results for $N=24$. 
Time evolution of the wave function $\Psi(\tau)$ is obtained by using the quasi eigenstates in the Lanczos algorithm~\cite{park, prelovsek} as 
\begin{align}
\ket{\Psi(\tau+\delta\tau)}  
=\sum_{j=1}^{M}e^{-{\rm i}\epsilon_{j}\delta\tau}\ket{\psi_{j}}\braket{\psi_{j}|\Psi(\tau)} ,
\label{e3}
\end{align}
where $\epsilon_{j}$ and $\ket{\psi_{j}}$ are the eigenvalues and
eigenstates, respectively, 
in the so-called order-$M$ Krylov subspace $\mathcal{K}_{M}(H,\ket{\Psi(\tau)})$,
and $\delta \tau$ is the time step.
We consider the case of $t=t'$, $V+V'=12t$; the number of the fermions is $N/2$. 
Energy and time parameter values are given as units of $t$ and $1/t$, respectively. 
We choose $M=15$ and $\delta \tau=0.005/t$ which are sufficient to obtain results with high enough accuracy. 
A damping factor is chosen to be $\tau_p = 3/t$, and 
the polarization direction is set parallel to the $y$ direction, i.e. $\bm{e} \parallel \bm{y}$. 
As for the results discussed in this Letter, we do not find any considerable difference to other polarization directions. 

Let us start from the results in the ground state before photoexcitation. 
In Fig.~\ref{fig:gs}(b), we show the charge correlation functions defined by 
$C(\bm{k}) = N^{-2}\sum_{i,j}
\braket{(n_{i}-1/2)(n_{j}-1/2)} 
e^{{\rm i}\bm{k}\cdot\bm{R}_{ij}}$ 
where the momentum ${\bm k}$ is defined in the Brillouin zone shown in Fig.~\ref{fig:gs}(a). 
The existence of three phases is confirmed; 
(i) $V/t < 5.5$ where $C(\pi/\sqrt{3}, \pi)$ and  $C(0, \pi)$ are dominant, 
(ii) $5.5 \le V/t \le 6.4 $ where $C(2\pi/\sqrt{3}, 2\pi/3)$ is dominant except for a narrow region around $V/t \sim 6.4$ where  $C(2\pi/\sqrt{3}, \pi/3)$ appears, 
and (iii) $6.4<V/t$ where $C(2\pi/\sqrt{3}, 0)$ is dominant. 
The correlations at $(0, \pi)$ and  at $(2\pi/\sqrt{3}, 0)$ are identified as the  h-stripe and v-stripe COs, respectively, shown in Fig.~\ref{fig:gs}(c), 
which are realized in the classical limit of 
weak fermion hopping with $V' >V$ and $V'<V$, respectively. 
The charge correlation at $(\pi/\sqrt{3}, \pi)$ observed in $V<5.5$ is attributed to the diagonal stripe-type (d-stripe) CO state, 
which is expected to vanish in the thermodynamic limit.~\cite{hotta} 
Around the frustration point, i.e. $V=V'$, the correlation at
$(2\pi/\sqrt{3}, 2\pi/3)$  implies the metallic three-fold CO realized
due to the fermion kinetic effect.~\cite{hotta, nishimoto}. 
Accordingly, we write in the following as 
${\bm k}_{\rm H}=(0, \pi)$, 
${\bm k}_{\rm V}=(2\pi/\sqrt{3}, 0)$, and ${\bm k}_{\rm 3}=(2\pi/\sqrt{3}, 2\pi/3)$. 

\begin{figure}[t]
\includegraphics[width=\columnwidth,clip]{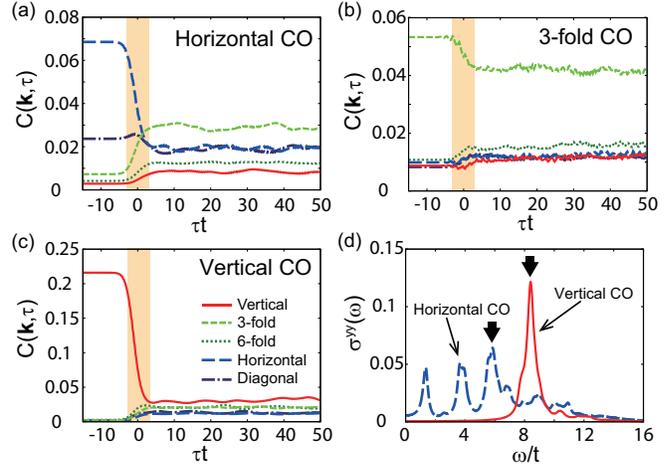}
\caption{
(Color online)
(a) Time-dependences of the charge correlation functions in the h-stripe ($V/t=5.4$), 
(b) three-fold ($V/t=6$), and (c) v-stripe ($V/t=6.5$) CO phases. 
Shaded areas represent the time intervals when the pump pulses are introduced.  
(d) The optical conductivity spectra before photo-irradiation in the h-stripe, and v-stripe CO phases. 
Bold arrows represent the energies which are chosen as the energies of the pump photons.
Pump photon amplitudes are chosen to be $A_{p}=2.4$, in which the absorbed photon numbers are approximately 1.7.  
}
\label{fig:ckt}
\end{figure}

We calculate the regular part of the optical conductivity defined by 
\begin{align}
\sigma^{\alpha \alpha}(\omega) = -\frac{1}{N\omega} \text{Im}
\bra{\Psi_{0}} j^{\alpha} \frac{1}{\omega-\cal{H}+E_{0}+{\rm i}\eta} j^{\alpha} \ket{\Psi_{0}},  \label{e6}
\end{align}
where $\Psi_0$ and $E_0$ are the ground-state wave function and energy, respectively, $j^{\alpha}$ is a current operator with a Cartesian coordinate $\alpha(= x, y)$, and $\eta$ is an infinitesimal constant.
Results in the h-stripe and v-stripe CO phases are shown in Fig~\ref{fig:ckt}(d). 
As shown in Fig.~\ref{fig:gs}(b), the adopted parameters correspond to vicinities of the phase boundaries.
In contrast to an almost single-peak structure in the v-stripe CO phase, 
multiple peaks are seen down to about $\omega/t=1$ in the h-stripe CO phase. 
The pump-photon energies used in the real-time simulations presented in Figs.~\ref{fig:ckt}(a) and (c) are marked by the bold arrows in Fig.~\ref{fig:ckt}(d).

Now, we show the transient charge structure. 
In Figs.~\ref{fig:ckt}(a)-(c), the time-dependences of the charge-correlation functions are presented. 
Pump photon amplitudes are chosen so that the absorbed photon numbers are approximately 1.7.
In all cases, the charge correlations at the momenta for the initial CO phases are  reduced  after the photo-excitations, implying the collapse of the initial COs. 
Compared with the insulating h-stripe and v-stripe phases, 
the initial CO correlation in the metallic three-fold CO phase remains to be large. 
Namely, the three-fold CO structure is insensitive to the photo-irradiation. 
In contrast, the photo-excitation to the h-stripe CO phase
gives rise to characteristic time-evolution. There,
not only the decrease in $C({\bm k}_{\rm H})$, but also the increase in $C({\bm k}_{\rm 3})$ is noticeably seen, and the largest correlation function is interchanged. 
%
On the other hand, such a transition is 
not seen in the v-stripe CO phase,
and $C(\bm{k})$ becomes more or less featureless. 

To examine the photo-induced charge structure in more detail, 
we introduce the change in the correlation functions per number of the absorbed photons defined by 
$\Delta C(\bm{k}) =  \left \{ C^{\text{\rm av}}(\bm{k}) - C^0(\bm{k} ) \right \} /N_p$ 
where $C^{\text{\rm av}}(\bm{k})$ and $C^0(\bm{k})$ are the correlation function  averaged during $15\le \tau t \le 50$ and that before photo-excitation, respectively.  
We define the number of the absorbed photons by 
$N_p=(E^{\text{\rm av}}-E_{0})/\omega_{p}$
where $E^\text{\rm av}$ is the energy expectation averaged during $15\le \tau t \le 50$, 
and $E_0$ is the energy before the photo-excitation. 
Pump photon amplitude is fixed to be $A_{p}=0.4$, and 
the photon energy  $\omega_{p}$ is chosen so that $N_p$ takes the maximum.
\begin{figure}[t]
\includegraphics[width=0.9\columnwidth,clip]{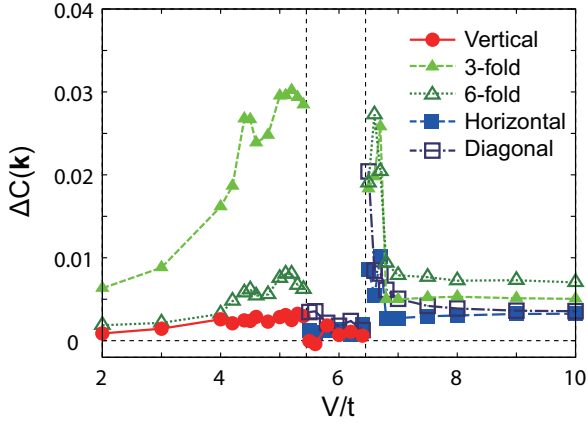}
\caption{
(Color online)
Changes of the charge correlation functions due to the photo-excitation normalized by the effective number of the absorbed photons. 
Filled circles, filled triangles, open triangles, filled squares and open squares represent 
$\Delta C({\bm k}_{\rm V})$ (v-stripe), $\Delta C({\bm k}_{\rm 3})$ (three-fold),  $\Delta C(2\pi/\sqrt{3}, \pi/3)$ (six-fold), $\Delta C({\bm k}_{\rm H})$ (h-stripe), and $\Delta C(\pi/\sqrt{3}, \pi)$ (d-stripe), respectively. 
The correlation functions at the momenta corresponding to the initial CO structures 
are not plotted. 
Vertical broken lines represent the phase boundaries. 
}
\label{fig:dck}
\end{figure}
Results 
are shown in Fig.~\ref{fig:dck} 
(the correlation functions corresponding to the initial CO structures are not plotted). 
It is clearly seen that in a wide region of the h-stripe CO phase ($V/t<5.5$), 
$\Delta C({\bm k}_{\rm 3})$  is larger than the other charge correlations, and increases toward the phase boundary. 
Although a similar critical-like behavior is seen in the vicinity of the phase boundary in the v-stripe CO phase ($V/t>6.4$), 
not only $\Delta C({\bm k}_{\rm 3})$ but also other correlations increase toward the boundary. 
This behavior implies that the v-stripe CO phase is not changed into a specific CO phase but to a featureless state. 
Only weak changes in $\Delta C(\bm{k})$ are seen in the three-fold CO phase ($5.5 \le V/t \le 6.5$). 

\begin{figure}[t]
\includegraphics[width=\columnwidth,clip]{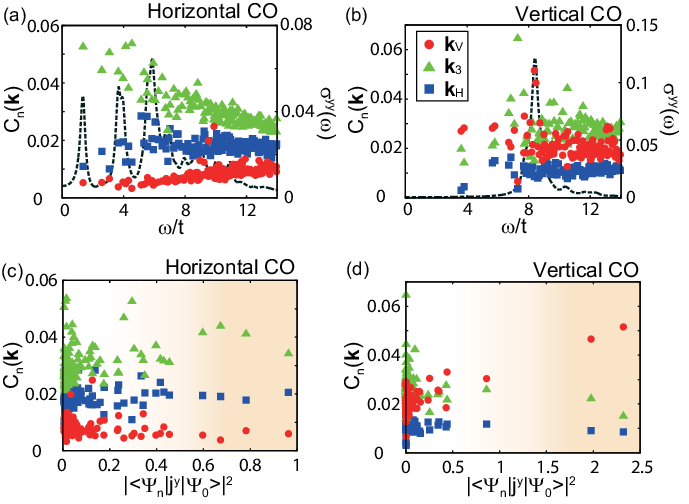}
\caption{
(Color online)
Charge correlation functions in the photo-excited eigenstates in the (a)
 h-stripe, and (b)v-stripe CO phases.
The optical conductivity spectra are also shown. 
(c) and (d): 
the charge correlation functions in the photo-excited states plotted as functions of the optical transition amplitude $|\braket{\Psi_{n}|j^{y}|\Psi_{0}}|^{2}$. 
Circles, triangles and squares 
represent  $C({\bm k}_{\rm V})$ (v-stripe),  $C({\bm k}_{\rm 3})$ (three fold), and $C({\bm k}_{\rm H})$ (h-stripe), respectively. 
Parameter values of $V$ in (a) and (c) [(b) and (d)] are the same with those in Fig.~\ref{fig:ckt}(a) [(c)]. 
}
\label{fig:exst}
\end{figure}
To identify the origin of the different photo-induced real-time dynamics in the h-stripe and v-stripe COs presented above, we examine the photo-excited eigenstates directly. 
The optical-allowed excited states $\ket{\Psi_{n}}$ with energy $E_n$ are given by the equation 
$\left\{ ({\cal{H}}-E_{n})^{2}+\eta^{2} \right\}\ket{\Psi_{n}} = - \eta j^{\alpha} \ket{\Psi_{0}}$ 
with an infinitesimal constant $\eta$. 
This is calculated by using the conjugated gradient method where $\eta$ is chosen to be $\eta/t=10^{-5}$. 
We calculate the charge correlation functions, $C_n({\bm k})$, in the optically-allowed excited states $\ket{\Psi_{n}}$. 
Results are shown in Fig.~\ref{fig:exst}(a) and (b) for the h-stripe and v-stripe CO phases, respectively, together with 
the optical conductivity spectra. 
The charge correlations at ${\bm k}_{\rm H}$, ${\bm k}_{\rm V}$, and ${\bm k}_{\rm 3}$ are shown.  
In the h-stripe CO phase (Fig.~\ref{fig:exst}(a)),
$C({\bm k}_{\rm 3})$ ($C({\bm k}_{\rm V})$) is the largest (smallest) among the three in almost all the optical-allowed excited states. 
This is consistent with the real-time photo-induced dynamics in the h-stripe CO phase shown in Fig.~\ref{fig:ckt} (a) and Fig.~\ref{fig:dck}. 
In the v-stripe CO phase (Fig.~\ref{fig:exst}(b)), on the other hand, 
the largest correlation functions in the excited states are not only 
$C({\bm k}_{\rm 3})$ but also $C(\bm{k}_{\rm V})$, 
in particular around the main optical absorption peak at $\omega/t
\simeq 8$.  
This different photo-excited states in the two COs are clearly seen in 
Fig.~\ref{fig:exst} (c) and (d) for the h-stripe and v-stripe CO phases,
respectively, where the correlation functions are replotted as a function of the optical transition amplitude $|\braket{\Psi_{n}|j^{\alpha}|\Psi_{0}}|^{2}$. 
Difference between the results in the two COs  is remarkable in the region of the high transition amplitudes; 
the largest correlation is 
 $C({\bm k}_{\rm 3})$ in the h-stripe CO phase and  is $C(\bm{k}_{\rm V})$ in the v-stripe CO. 
 
\begin{figure}[t]
\includegraphics[width=\columnwidth,clip]{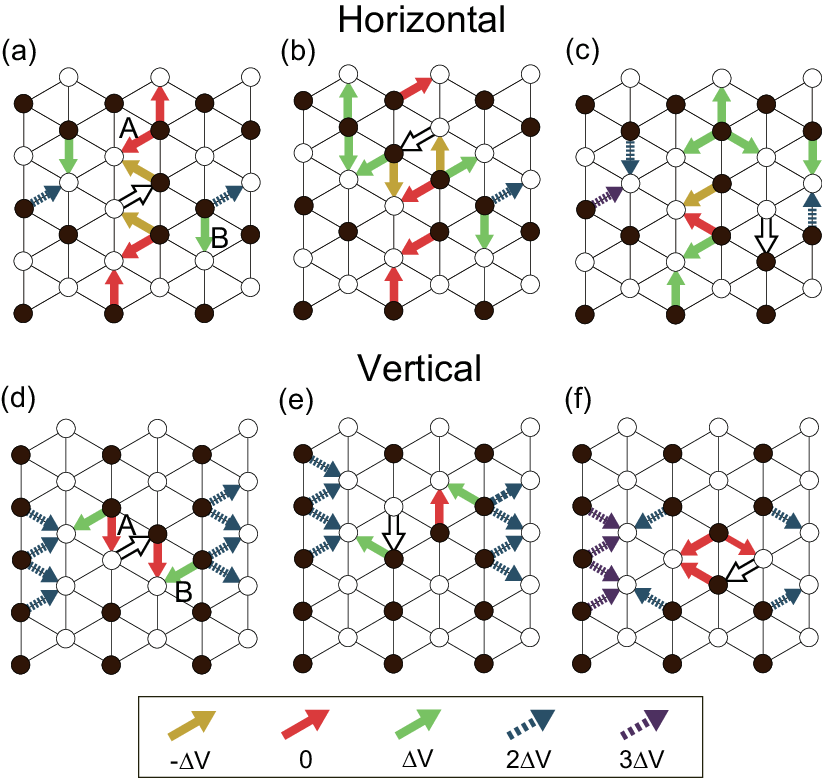}
\caption{
(Color online)
Schematic charge configurations of low-energy charge excitations in the photo-excited states. 
(a)-(c) are for the h-stripe CO phase, and (d)-(f) are for the v-stripe CO phase. 
Filled and open circles represent charge rich and poor sites, respectively. 
(a) a charge configuration where a fermion at a certain site hops (an open arrow) from the initial h-stripe CO phase. 
Filled arrows represent possible low-energy excitations due to the NN charge hoppings. 
(b) [(c)] represents a charge configuration in the case where a fermion hopping  indicated by the arrow A [B] in (a) occurs.
Filled and dashed arrows represent  possible low-energy excitations.
(d)-(f) in the v-stripe CO phase correspond to (a)-(c) in the h-stripe phase,  respectively. 
}
\label{fig:diss}
\end{figure}
Here, we give an interpretation for the numerical results above, and discuss 
frustration effects in the photo-excited state,
from the viewpoint in the strong coupling limit of $t \ll V \sim V'$. 
Figures~\ref{fig:diss}(a)-(c) and Figs.~\ref{fig:diss}(d)-(f) schematically show the charge configurations 
for the photo-excited states 
in the h-stripe and v-stripe CO phases, respectively. 
Let us start from Fig.~\ref{fig:diss}(a) where 
a charge at a certain site in the h-stripe CO phase hops to one of the NN sites (see the open arrow) 
due to the photo-irradiation. 
Filled and dashed arrows indicate possible low energy excitations in this photo-excited state 
through a NN charge hopping within the excitation energy $3|V'-V|$. 
Then, we choose the cases where one of these charge hoppings, A or B in Fig.~\ref{fig:diss}(a), is realized, 
changing the charge configuration to the ones shown in Fig.~\ref{fig:diss}(b) and (c), respectively. 
Further low energy excitations are induced due to NN charge hopping indicated by filled and dashed arrows. 
Existence of this avalanche-like processes implies that a local excitation leads to a metallic state which is energetically close to the h-stripe state,
i.e. the photo-induced 3-fold metallic state from the h-stripe CO phase.
This characteristics originate from the facts that i) a number of almost degenerate states exist 
in the photo-excited state near the frustration point, $V \sim V'$, and ii) excitations to these states are optically allowed from the initial h-stripe CO. 
As for the v-stripe CO ground state, 
analogously generated charge configurations in the photo-excited state and possible low-energy excitation processes are presented in Figs.~\ref{fig:diss}(d)-(e). 
In comparison with the h-stripe CO phase, 
the possible low-energy excitation processes are limited. In particular, the number of the excitations within the energy $|V'-V|$ are much less.
This fact implies that the v-stripe CO phase is robust under the photo-excitations; except for the region at very vicinity of the frustration point, i.e. $V=V'$,
charge excitations, which are optically allowed from the v-stripe CO, are not extended, and then
the initial v-stripe CO correlation is merely weakened by photo-irradiation. 

In summary, the photo-excitation dynamics in the interacting charge-ordered system on a geometrically frustrated triangular lattice are examined.
A spinless fermion model with fermion hoppings and Coulomb interactions between the NN sites is  analyzed by the numerical exact diagonalization method. 
Photo-excited states in the two canonical COs, the h-stripe and v-stripe COs, are focused on. 
Real time dynamics, as well as the charge structures in the excited eigenstates, are distinct in the two COs; 
the photo-induced transition from the h-stripe CO to the three-fold CO occurs. 
On the other hand, 
only weakening of CO due to photo-irradiation is seen in the v-stripe CO phase. 
This difference is attributed to the fact that there is a number of
low-energy excited states which are optically allowed from the h-stripe
CO, in contrast to the case of the v-stripe CO. 
This is a manifestation of the frustration effect in the photo-excited
state, whose concept is expected to be applied to a wide class of the frustrated electron systems.

We thank M.~Naka, J.~Nasu, K.~Iwano, and S.~Iwai for their helpful discussions.
This work was supported by 
JSPS KAKENHI Grant Numbers 26400377 and 26287070,
and the RIKEN iTHES project.
Some of the numerical calculations were performed using the supercomputing facilities at ISSP, the University of Tokyo.


\begin{thebibliography}{9}

\bibitem{JPSJ_frustration}
{\it Special Topics: Novel States of Matter Induced by Frustration}, 
J. Phys. Soc. Jpn. {\bf 79}, 001101-001112 (2010). 

\bibitem{book_frustration}
{\it  Frustrated Spin Systems}, 
edited by H T Diep, 
(World Scientific, New Jersey, 2004). 

\bibitem{shimizu}
Y. Shimizu, K. Miyagawa, K. Kanoda, M. Maesato, and G. Saito,
Phys. Rev. Lett. {\bf 91}, 107001 (2003). 
\bibitem{han}
T.-H. Han, J.S. Helton, S. Chu, D.G. Nocera, J.A. Rodriguez-Rivera, C. Broholm, and Y.S. Lee,
Nature {\bf 492}, 406 (2012).

\bibitem{harris}
M.J. Harris, S.T. Bramwell, D.F. McMorrow, T. Zeiske, and K.W. Godfrey,
Phys. Rev. Lett. {\bf 79}, 2554 (1997). 
\bibitem{nakatsuji}
S. Nakatsuji, Y. Nambu, H. Tonomura, O. Sakai, S. Jonas, C. Broholm, H. Tsunetsugu, Y. Qiu, and Y. Maeno,
Science {\bf 309}, 1697 (2005).
\bibitem{kimura}
T. Kimura, T. Goto, H. Shintani, K. Ishizaka, T. Arima, and Y. Tokura,
Nature {\bf 426}, 55 (2003).
\bibitem{anderson}
P. W. Anderson,
Phys. Rev. {\bf 102}, 1008 (1956).
\bibitem{verwey}
E. J. W. Verwey, and P. W. Haaymann, 
Physica {\bf 8}, 979 (1941).
%
%
\bibitem{seo} 
H. Seo, J. Merino, H. Yoshioka, and M. Ogata,  J. Phys. Soc. Jpn. {\bf 75}, 051009 (2006). 

\bibitem{merino} 
J. Merino, H. Seo and M. Ogata, Phys. Rev. B {\bf 71}, 125111 (2005).

\bibitem{hotta}
C. Hotta, N. Furukawa, A. Nakagawa, and K. Kubo,
J. Phys. Soc. Jpn. {\bf  75}, 123704 (2006). 

\bibitem{nishimoto0}
S. Nishimoto, M. Shingai, and Y. Ohta, 
Phys. Rev. B, {\bf 78}, 035113 (2008). 
%
%
%
\bibitem{nasu}
{\it Photo-Induced Phase Transitions}, 
K. Nasu, (World Scientific, New Jersey, 2004). 
\bibitem{JPSJ_PIPT}
{\it Special Topics: Photo-Induced Phase Transitions and their Dynamics}, 
J. Phys. Soc. Jpn. {\bf 75}, 001101-001108 (2006).

\bibitem{chollet}
M. Chollet, L. Guerin,  N. Uchida,  S. Fukaya,  H. Shimoda,     T. Ishikawa,  K. Matsuda, T. Hasegawa,   A. Ota,  H. Yamochi,   G. Saito,  R. Tazaki,  S. Adachi,  S. Koshihara, 
Science {\bf 307}, 86 (2005). 

\bibitem{iwai}
S. Iwai, K. Yamamoto, A. Kashiwazaki, F. Hiramatsu, H. Nakaya, Y. Kawakami, K. Yakushi, H. Okamoto, H. Mori, and Y. Nishio,  
Phys. Rev. Lett. {\bf 98}, 097402 (2007).

\bibitem{rohwer}
T. Rohwer, S. Hellmann,	M. Wiesenmayer,	C. Sohrt, A. Stange, B. Slomski, A. Carr, Y. Liu,	L. M. Avila, M. Kallane, S. Mathias, L. Kipp, K. Rossnagel, M. Bauer
Nature, {\bf   471}, 490 (2011). 
    
\bibitem{stojchevska}
L. Stojchevska, I. Vaskivskyi, T. Mertelj, P. Kusar, D. Svetin, S. Brazovskii, and D. Mihailovic, 
Science {\bf 344}, 177 (2014). 

\bibitem{yonemitsu}
K. Yonemitsu and N. Maeshima, 
Phys. Rev. B {\bf 76}, 075105 (2007). 

\bibitem{iwano}
K. Iwano, 
Phys. Rev. Lett. {\bf 102}, 106405 (2009). 

\bibitem{kanamori}
Y. Kanamori, H. Matsueda and S. Ishihara, 
Phys. Rev. Lett. {\bf 103}, 267401 (2009). 

\bibitem{gomi}
T. Tatsumi, H. Gomi, A. Takahashi, Y. Hirao, and M. Aihara, 
J. Phys. Soc. Jpn. {\bf 81}, 034712 (2012). 

\bibitem{naka}
M. Naka, and H. Seo,
J. Phys. Soc. Jpn. {\bf 83}, 053706 (2014).

\bibitem{nishimoto}
S. Nishimoto, and C. Hotta,
Phys. Rev. B {\bf 79}, 195124 (2009).

\bibitem{park}
T. J. Park, and J. C. Light,
J. Chem. Phys. {\bf 85}, 5870 (1986).

\bibitem{prelovsek}
P. Prelovsek, and J. Bonca,
arXiv:1111.5931.


\end{thebibliography}
\end{document}